\newcommand{\abs}[1]{\left\vert#1\right\vert}

\newcommand{\ket}[1]{\left\vert #1\right\rangle}

\newcommand{\bra}[1]{\left\langle #1\right\vert}

\newcommand{\brkt}[2]{\left\langle #1 \vert #2\right\rangle}

\newcommand{\braket}[3]{\left\langle #1 \right\vert #2\left\vert #3\right\rangle}

\newcommand{\D}{\displaystyle}

\newcommand{\SSt}{\scriptscriptstyle}

\documentclass[preprint,aps,prc,showpacs,amsmath,amssymb]{revtex4-1}
\usepackage{times}          
\usepackage{graphicx}
\usepackage{bm}
\usepackage{amsmath}
\usepackage{amsxtra}
\usepackage{amstext}
\usepackage{amssymb}
\usepackage{latexsym}
\usepackage{dsfont}

\begin{document}
\title{{\large  $\mathbf{\Omega^{-}}$, $\mathbf{\Xi^{*\,-}}$, $\mathbf{\Sigma^{*\,-}}$ and $\mathbf{\Delta^{-}}$ decuplet baryon electric charge form factor $\mathbf{F_{1}(q^{2})}$}}
\author{Milton Dean Slaughter}
\address{Department of Physics, Florida International University, Miami, Florida 33199, USA}
\email{slaughtm@FIU.Edu,\;Slaughts@PhysicsResearch.Net}

\begin{abstract}
The magnetic moment---a function of the electric charge form factor $F_{1}(q^{2})$ and the magnetic dipole form factor $F_{2}(q^{2})$ at zero four-momentum transfer $q^{2}$---of the ground-state $U$-spin $=\frac{3}{2}$ baryon decuplet $\Delta^{-}$, $\Xi^{*\,-}$, $\Sigma^{*\,-}$ and $\Omega^{-}$ and their ground-state spin $\frac{1}{2}$ cousins $p$, $n$, $\Lambda$, $\Sigma^{+}$, $\Sigma^{0}$, $\Sigma^{-}$, $\Xi^{+}$, and $\Xi^{-}$ have been studied for many years with a modicum of success---only the magnetic moment of the $\Omega^{-}$ has been accurately determined. In a recent study by us utilizing the infinite momentum frame, we calculated the magnetic moments of the \emph{physical} decuplet $U$-Spin $=\frac{3}{2}$ quartet members in terms of that of the $\Omega^{-}$ without ascribing any specific form to their quark structure or intra-quark interactions. That study determined $F_{2}(q^{2})$ and was conducted nonperturbatively where the decuplet baryon momenta were all collinear.  In this follow-up research---again utilizing the infinite momentum frame but now allowing for non-collinear momenta---we are able to determine $F_{1}(q^{2})$ where $q^{2}\leq 0$.  We relate the electric charge form factor $F_{1}(q^{2})$ of the \emph{physical} decuplet $S\neq -3$, $U$-spin $=\frac{3}{2}$ quartet members to that of the $\Omega^{-}$ ($S=-3$).
\end{abstract}
\maketitle

The properties of the ground-state $U$-spin $=\frac{3}{2}$ baryon decuplet magnetic moments $\Delta^{-}$, $\Xi^{*\,-}$, $\Sigma^{*\,-}$ and $\Omega^{-}$ along with their ground-state spin-$\frac{1}{2}$ cousins $p$, $n$, $\Lambda$, $\Sigma^{+}$, $\Sigma^{0}$, $\Sigma^{-}$, $\Xi^{+}$, and $\Xi^{-}$ have been studied for many years with a modicum of success.  Although the masses (pole or otherwise) and decay aspects and other physical observables of some of these particles have been ascertained, the magnetic moments of many are yet to be determined.  For the spin $=\frac{3}{2}$ baryon decuplet, the experimental situation is poor---from the Particle Data Group \cite{Nakamura:2010zzi}, only the magnetic moment of the $\Omega^{-}$ \cite{Wallace:1995pf} has been accurately determined. The magnetic moment is a function of the electric charge form factor $F_{1}(q^{2})$ and the magnetic dipole form factor $F_{2}(q^{2})$ at zero four-momentum transfer $q^{2}\equiv -Q^{2}$. The reasons for this paucity of data for the decuplet particle members are the very short lifetimes owing to available strong interaction decay channels and the existence of nearby particles with quantum numbers that allow for configuration mixing which greatly increases the difficulty of experimental determination of physical observables. The $\Omega^{-}$ (strangeness $S=-3$) is an exception in that it is composed of three valence $s$ quarks that make its lifetime substantially longer (weak interaction decay) than any of its decuplet partners, which have many more decay channels available.  Even for the $\Omega^{-}$, away from the static ($q^{2}= 0$) limit, the electric charge and magnetic dipole form factors are not known.

A number of theoretical models have been put forth over the past few decades. In addition to the simplest $SU(3)$ model, seminal ones are the $SU(6)$ models put forth by Beg \emph{et al.} \cite{Beg:1964nm} and Gerasimov \cite{Gerasimov:1966}.  An excellent source of information on the aforementioned topics, references, and other seminal models is the book by Lichtenberg \cite{Lichtenberg:1978pc}.  Typically, these models invoke the additivity hypothesis where a hadron magnetic moment is given by the sum of its constituent quark magnetic moments.

In a recent research publication \cite{Slaughter:2011??} by us utilizing the infinite momentum frame in conjunction with the fact that the four-vector electromagnetic current $j^{\mu}_{em}$ obeys the equal time commutator $\left[V_{K^{0}}, j^{\mu}_{em}\right]=0$ which is valid even in the presence of symmetry breaking, we illustrated how one may calculate the magnetic moments of the \emph{physical} decuplet $U$-Spin $=\frac{3}{2}$ quartet members in terms of that of the $\Omega^{-}$ without ascribing any specific form to their quark structure or intra-quark interactions \cite{Slaughter:2011??,Oneda:1970ny,Oneda:1985wf,Slaughter:1988hx,Oneda:1989ik,alfaro63}.  In that nonperturbative study, the magnetic dipole form factor $F_{2}(q^{2})$ was determined and was conducted where the decuplet baryon momenta were all collinear.

Over the years, a number of theoretical and computational investigations involving the magnetic moments of the $\Omega^{-}$ and the $\Delta^{-}$ and lattice quantum chromodynamics (LQCD) (quenched and unquenched, unphysical pion mass) techniques have been used with some progress.  In particular, some LQCD approaches show promise \cite{Boinepalli:2009sq,Aubin:2009qp,Aubin:2010jc}.  A review that focuses on some theoretical and experimental approaches to the study of specific processes involving the $\Delta(1232)$ can be found in Ref.~\cite{Pascalutsa:2006up}.  It is very important to realize that knowledge of the behavior of the decuplet form factors (or corresponding multipole moments) is critical to our understanding of QCD---standard model, enhanced standard model, lattice gauge models, superstring models, or entirely new models--- since for these models to have relevance they must be capable of yielding already known results at low or medium energy.

In this work the infinite momentum frame---in conjunction with the fact that the four-vector electromagnetic current $j^{\mu}_{em}$ obeys the equal time commutator $\left[V_{K^{0}}, j^{\mu}_{em}\right]=0$ even in the presence of symmetry breaking---is used to study the electric charge form factor $F_{1}(q^{2})$ of the \emph{physical} spin $=\frac{3}{2}$ baryon decuplet $U$-spin $=\frac{3}{2}$ quartet members without ascribing any specific form to their quark structure or intra-quark interactions.  We again utilize the infinite momentum frame but now allow for non-collinear momenta allowing us to determine $F_{1}(q^{2})$ where $q^{2}\leq 0$.  We relate the electric charge form factor $F_{1}(q^{2})$ of the \emph{physical} decuplet $S\neq -3$, $U$-spin $=\frac{3}{2}$ quartet members to that of the $\Omega^{-}$ ($S=-3$).

As before \cite{Slaughter:2011??} all equal-time commutation relations (ETCRs) involve at most one current density, ensuring that problems associated with Schwinger terms are avoided.   ETCRs involve the vector and axial-vector charge generators (the
 $V_{\alpha }$ and $A_{\alpha }$ $\{\alpha
=\pi ,K,D,F,B,\ldots .\}$) of the symmetry groups of QCD and they
are valid even though these symmetries are broken
\cite{Slaughter:2011??,Oneda:1970ny,Oneda:1985wf,Slaughter:1988hx,Oneda:1989ik,alfaro63,gell-mann63,adler,Weisberger} and {\em even when the \mbox{Lagrangian} is
not known or cannot be constructed}.

A fundamental part of the dynamical concept of asymptotic
$SU_{F}(N)$ symmetry in the infinite momentum frame \cite{Slaughter:2011??,Oneda:1970ny,Oneda:1985wf,Slaughter:1988hx,Oneda:1989ik} is the behavior of the vector charge
$V_{\alpha }$ when acting on a physical state which has momentum
$\vec{k}$ ($|\vec{k}|\rightarrow \infty $), helicity $\lambda $,
and $SU_{F}(N)$ index $\alpha $: The physical annihilation
operator $a_{\alpha }(\vec{k},\lambda )$ of a physical
on-mass-shell hadron maintains its linearity (including asymptotic $%
SU_{F}(N)$ particle mixings) under flavor transformations
generated by the charge $V_{\alpha }$ but only in the limit
$|\vec{k}|\rightarrow \infty $.

It is in the $\infty
$-momentum frame where one finds that the {\em {physical}} annihilation operator
$a_{\alpha }(\vec{k},\lambda )$ is related {\em {linearly}} to the
{\em {representation}} annihilation operator
$a_{j}(\vec{k},\lambda )$. In contrast to the representation states denoted by $\vert j, \vec k,
\lambda\rangle$ that belong to irreducible representations, the physical states denoted by $\vert \alpha,\vec k, \lambda\rangle$ do not.  Rather, they are linear combinations of representation states plus non-linear corrective terms that are best calculated in a frame where mass differences are deemphasized such as the $\infty
$-momentum frame.  When flavor symmetry is exact, which Lorentz frame one uses to analyze current-algebraic sum rules does not matter and is a matter of taste and convenience of calculation, whereas, when one deals with current-algebraic sum rules in broken symmetry, the choice of frame takes on paramount importance because one wishes to emphasize the calculation of leading order contributions while simultaneously simplifying the calculation of symmetry breaking corrections.  The $\infty
$-momentum frame is especially suited for broken symmetry calculations because mass differences are kinematically suppressed \cite{Slaughter:2011??,Oneda:1970ny,Oneda:1985wf,Slaughter:1988hx,Oneda:1989ik,alfaro63}.

The physical vector charge $V_{K^{0}}$ may be written as $V_{K^{0}}=V_{6}+iV_{7}$ and the physical electromagnetic current $j_{em}^{\mu}(0)$ may be written ($u$, $d$%
, $s$, $c$, $b$, $t$ quark system) as $j_{em}^{\mu}(0) =V_{3}^{\mu}(0)+(1/
\sqrt{3})V_{8}^{\mu}(0)-(2/3)^{1/2}V_{15}^{\mu}(0)+(2/5)^{1/2}V_{24}^{\mu}(0)- (3/5)^{1/2}V_{35}^{\mu}(0)
+(1\sqrt{3})V_{0}^{\mu}(0)$.  One may verify that the commutation relation $\left[V_{K^{0}},j^{\mu}_{em}(0)\right]=0$ holds (\emph{i.e.}, the electromagnetic current is a $U$-spin singlet).

For the on-mass shell $J^{P}=3/2^{+}$ ground state decuplet baryon B with mass $m_{B}$, the Lorentz- covariant and gauge-invariant electromagnetic current matrix element in momentum space where the four-momentum vectors $P\equiv p_{1}+p_{2},q\equiv p_{2}-p_{1}$ and $\lambda_{1}$ and $\lambda_{2}$ represent helicity is given by:
\begin{equation}
\left\langle B{(}p_{2},\lambda _{2})\right| j_{em}^{\mu}(0)\left|
B(p_{1}{,\lambda _{1})}\right\rangle ={\frac{e}{{(2\pi )^{3}}}}\sqrt{{%
\frac{{m}_{B}^{2}}{{E_{B}^{t}E_{B}^{s}}}}}\bar{u}^{\alpha}_{B}(p_{2},\lambda _{2})\left[ {%
\Gamma^{\mu }_{\alpha \beta}}\right] u^{\beta}_{B}\left( p_{1},\lambda _{1}\right),  \label{matrixeqn}
\end{equation}
\begin{equation}
\begin{array}{ccl}
\Gamma^{\mu }_{\alpha \beta} & = & g_{\alpha \beta}  \left\{  F_{1}^{B}(q^{2})\gamma^{\mu
}+\D{\frac{F_{2}^{B}(q^{2})i\sigma^{\mu \nu}}{2 m_{B}}  q_{\nu
}}   \right\}\\
& + & \D{\frac{q_{\alpha}q_{\beta}}{2 m_{B}^{2}}} \left\{ F_{3}^{B}(q^{2})\gamma^{\mu
}+\D{\frac{F_{4}^{B}(q^{2})i\sigma^{\mu \nu}}{2 m_{B}}  q_{\nu
}}   \right\},%
\end{array}
\label{matrixcoveqn}
\end{equation}
\noindent where $e=+\sqrt{4\pi\alpha}$, $\alpha=$ the fine structure constant, the $F_{i}^{B}$ are the four $\gamma^{*}BB$ form \mbox{factors}
[$F_{1}^{B}(0)\sim \mbox{electric charge in units of $e$}$, $(F_{1}^{B}(0)+F_{2}^{B}(0))\sim $ magnetic dipole moment in units of $e/(2m_{B})$] and $\Gamma^{\mu }_{\alpha \beta}$ is written in standard form \cite{Korner:1976hv}.
The electric charge multipole \mbox{amplitude} $G_{E}^{B}(q^{2})=[F_{1}^{B}(q^{2}) (3 - 2 \eta) + \eta \{F_{2}^{B}(q^{2}) (3 - 2 \eta) -
    2 (-1 + \eta) (F_{3}^{B}(q^{2}) + \eta\,F_{4}^{B}(q^{2}) )\}]/3$ [units of $e$],
the magnetic dipole multipole amplitude $G_{M}^{B}(q^{2})=[ (5 - 4 \eta) ( F_{1}^{B}(q^{2})+F_{2}^{B}(q^{2}))    -
 4\,\eta \,(-1 + \eta) \, (F_{3}^{B}(q^{2}) + F_{4}^{B}(q^{2}))  ]/5$ [units of $e/(2m_{B})$],
  the electric quadrupole multipole \mbox{amplitude} $G_{Q}^{B}(q^{2})=F_{1}^{B}(q^{2})+F_{3}^{B}(q^{2})\, (-1 + \eta)\, +\eta \, \{F_{2}^{B}(q^{2})+F_{4}^{B}(q^{2})\, (-1 + \eta)\}$ [units of $e/m_{B}^2$], and the magnetic octupole multipole amplitude $G_{O}^{B}(q^{2})=[F_{1}^{B}(q^{2})+F_{2}^{B}(q^{2})+(-1+\eta)\{F_{3}^{B}(q^{2}) +F_{4}^{B}(q^{2})\} ]\sqrt{6}$ [units of $e/(2m_{B}^3)$] where $\eta \equiv q^{2}/(4m_{B}^2)$.   $Q_{B}=$ charge of baryon $B$ in units of $e$, $\mu_{B}$ is the magnetic moment (measured in nuclear magneton units $\mu_{N}=e/(2 m)$, $m=$ proton mass) of baryon $B$ and explicitly:

\begin{eqnarray}
  &&F_{1}^{B}(0)\,e  = Q_{B}\,,\\
  &&\mu_{B} = \left\{[F_{1}^{B}(0)+F_{2}^{B}(0)] (\frac{m}{m_{B}})\right\} \mu_{N} .
\end{eqnarray}

The baryon Rarita-Schwinger \cite{Rarita:1941mf} spinor $u_{B}^{\mu}(\nu_{B},\theta,\lambda)$ with helicity $\lambda$, three-momentum $\vec{p}$ with angle $\theta$ referred to the $\hat{z}$-axis, energy $E_{B}^{p}$, and velocity parameter $\nu_{B}=\sinh^{-1}({\abs{\vec p \,}}/m_B)$ is given by:
\begin{equation}
u_{B}^{\mu}(\nu_{B},\theta,\lambda)=
\sum_{m_1=-\frac{1}{2}}^{+\frac{1}{2}}\sum_{m_2=-1}^{+1}
\brkt{1/2, 1, 3/2}{m_1, m_2, \lambda}
u_{B}(\nu_{B},\theta,m_1)\epsilon_{B}^{\mu}(\nu_{B},\theta,m_2),
\label{rsspinoreqn}
\end{equation}
\begin{eqnarray}
u_{B}(\nu_{B},\theta,m_1)&=&
\left(
\begin{array}{l}
 \cosh(\frac{\nu_{B}}{2}) [ \cos(\frac{\theta}{2})\,\delta_{m_1,\,\frac{1}{2}} -\sin(\frac{\theta}{2})\,\delta_{m_1,\,-\frac{1}{2}} ]\\
 \cosh(\frac{\nu_{B}}{2}) [ \sin(\frac{\theta}{2})\,\delta_{m_1,\,\frac{1}{2}} +\cos(\frac{\theta}{2})\,\delta_{m_1,\,-\frac{1}{2}} ]  \\
 \sinh(\frac{\nu_{B}}{2}) [ \cos(\frac{\theta}{2})\,\delta_{m_1,\,\frac{1}{2}} +\sin(\frac{\theta}{2})\,\delta_{m_1,\,-\frac{1}{2}} ] \\
 \sinh(\frac{\nu_{B}}{2}) [ \sin(\frac{\theta}{2})\,\delta_{m_1,\,\frac{1}{2}} -\cos(\frac{\theta}{2})\,\delta_{m_1,\,-\frac{1}{2}} ]
 \end{array}
\right), \label{Bspinoreqn} \\ \nonumber
\end{eqnarray}
\begin{eqnarray}
\epsilon_{B}^{\mu}(\nu_{B},\theta,m_2)&=&
\left(
\begin{array}{l}
 \sinh(\nu_{B})\, \delta_{m_2,\,0}  \\
 -\frac{m_2}{\sqrt{2}}\cos(\theta)\,\delta_{\abs{m_2},\,1}+\cosh(\nu_{B}) \sin(\theta)\,\delta_{m_2,\,0}  \\
 -\frac{i}{\sqrt{2}} \,\delta_{\abs{m_2},\,1} \\
 \frac{m_2}{\sqrt{2}}\sin(\theta)\,\delta_{\abs{m_2},\,1}+\cosh(\nu_{B}) \cos(\theta)\,\delta_{m_2,\,0}
 \end{array}
\right). \label{polveceqn} \\ \nonumber
\end{eqnarray}
$\epsilon_{B}^{\mu}(\nu_{B},\theta,m_2)$ is the baryon polarization ($m_{2}$) four-vector, $u_{B}(\nu_{B},\theta,m_1)$ is a Dirac spinor with helicity index $m_{2}$, and $\brkt{1/2, 1, 3/2}{m_1, m_2, \lambda}$ is a Clebsh-Gordan coefficient where our conventions are those of Rose \cite{rose1957:etam}.  Physical states are normalized with $\brkt{\vec{p\,'}}{\vec{p}}=\delta^{3}%
(\vec{p\,'}\,-\vec{p})$ and Dirac spinors
are normalized by $\bar{u}^{(r)}(p)u^{(s)}(p)=\delta_{rs}$.  Our conventions for
Dirac matrices are $\left\{  \gamma^{\mu},\gamma^{\nu}\right\}
=2g^{\mu\nu}$ with $\gamma _{5}\equiv
i\gamma^{0}\gamma^{1}\gamma^{2}\gamma^{3}$, where $g^{\mu\nu}=$
Diag $(1,-1,-1,-1)$ \cite{Slaughter:2008zd}. The Ricci-Levi-Civita tensor is defined by
$\varepsilon _{0123}=-\varepsilon^{0123}=1=\varepsilon_{123}$.  As usual, we use natural units where $\hbar=c=1$.

Associated with baryon B are the four-momentum vectors $p_1$ (three-momentum $\vec{t}$ ($\vec{t}=t_z \hat{z}$), energy $E_{B}^{t}$) and $p_2$ (three-momentum $\vec{s}$ at angle $\theta$ ($0\leq\theta < \pi/2$) with the $\hat{z}$ axis, energy $E_{B}^{s}$) and we write:
\begin{eqnarray}
p_{1}^{\sigma}=t^{\sigma}=\left(
\begin{array}{c}
 m_{B}\cosh(\alpha_{B})\\
 0 \\
 0 \\
 m_{B}\sinh(\alpha_{B})
\end{array}
\right)=\left(
\begin{array}{c}
E_{B}^{t}\\
\vec{t}\\
\end{array}
\right)\label{momvecp1eqn}
,\\
p_{2}^{\sigma}=s^{\sigma}=\left(
\begin{array}{c}
 m_{B}\cosh(\beta_{B})\\
 m_{B}\sin(\theta) \sinh(\beta_{B})\\
 0 \\
 m_{B}\cos(\theta) \sinh(\beta_{B})
\end{array}
\right)=\left(
\begin{array}{c}
E_{B}^{s}\\
\vec{s}\\
\end{array}
\right).
\label{momvecp2eqn}
\end{eqnarray}

In Eqs.~(\ref{momvecp1eqn}) and (\ref{momvecp2eqn}), we take $s_{z}=r t_{z}$, where $r\, (\mbox{constant})\,\geq 1$.  In addition to obeying the Dirac equation---thus making the Gordon identities very useful--- the Rarita-Schwinger spinors satisfy the subsidiary conditions $\gamma_{\mu}u^{\mu}_{B}\left( p,\lambda\right)=p_{\mu}u^{\mu}_{B}\left( p,\lambda\right)=0$.

Previously \cite{Slaughter:2011??}, we utilized the commutator  $\left[V_{K^{0}},j^{\mu}_{em}(0)\right]=0$ inserted between the baryon pairs ($\bra{\Xi^{*\,-}s^{\sigma}}$,$\ket{\Omega^{-}t^{\sigma}}$),
($\bra{\Sigma^{*\,-}s^{\sigma}}$,$\ket{\Xi^{*\,-}t^{\sigma}}$), and
($\bra{\Delta^{-}s^{\sigma}}$,$\ket{\Sigma^{*\,-}t^{\sigma}}$) in the infinite momentum frame where each baryon had $Q_{B}=-e$, helicity $+3/2$ and $t_{z}\rightarrow \infty$ and $s_{z}\rightarrow \infty$ and $s_{x}=0$ (collinear, \emph{i.e.} $\vec{s}\times\vec{t}=0$). $\Gamma^{\mu }_{\alpha \beta}$ was written so that the $F_{2}^{B}$ form factor term contained the factor $P^{\mu}\equiv s^{\mu}+t^{\mu}$ which dominated other $\Gamma^{\mu }_{\alpha \beta}$ terms when $\mu=0$ in the $\infty$-momentum frame, thus serving as a $F_{2}^{B}$ projector for the matrix element $\braket{Bs^{\sigma},\lambda}{j_{em}^{\mu}(0)}{Bt^{\sigma},\lambda}$. In this research, we write $\Gamma^{\mu }_{\alpha \beta}$ in standard form \cite{Korner:1976hv} where the $F_{2}^{B}$ form factor term is not enhanced by the presence of $P^{\mu}$. As before \cite{Slaughter:2011??},
we utilize the commutator  $\left[V_{K^{0}},j^{\mu}_{em}(0)\right]=0$ inserted between the baryon pairs ($\bra{\Xi^{*\,-}s^{\sigma}}$,$\ket{\Omega^{-}t^{\sigma}}$),
($\bra{\Sigma^{*\,-}s^{\sigma}}$,$\ket{\Xi^{*\,-}t^{\sigma}}$), and
($\bra{\Delta^{-}s^{\sigma}}$,$\ket{\Sigma^{*\,-}t^{\sigma}}$) in the infinite momentum frame. The internal intermediate states saturating the commutator belong to the ground state decuplet baryons with helicity $+3/2$ which had the effect of restricting greatly the number of possible configuration mixing contributions coming from $56$ or spin $3/2$ members of $70$ excited states and other low-lying supermultiplets.  We have:
\begin{eqnarray}
\lefteqn{ \braket{\Xi^{*^-}s^{\sigma}}{V_{K^{0}}}{\Omega^{-}s^{\sigma}}    \braket{\Omega^{-}s^{\sigma}} {j_{em}^{\mu}} {\Omega^{-}t^{\sigma}} }  \nonumber \\
& & - \braket{\Xi^{*^-}s^{\sigma}}{j_{em}^{\mu}}{\Xi^{*^-}t^{\sigma}}   \braket{\Xi^{*^-}t^{\sigma}} {V_{K^{0}}} {\Omega^{-}t^{\sigma}} =0,\label{CRK1eqn}\\
\lefteqn{ \braket{\Sigma^{*\,-}s^{\sigma}}{V_{K^{0}}}{\Xi^{*^-}s^{\sigma}}   \braket{\Xi^{*^-}s^{\sigma}} {j_{em}^{\mu}} {\Xi^{*^-}t^{\sigma}} }  \nonumber \\
& & - \braket{\Sigma^{*^-}s^{\sigma}}{j_{em}^{\mu}}{\Sigma^{*^-}t^{\sigma}}   \braket{\Sigma^{*^-}t^{\sigma}} {V_{K^{0}}} {\Xi^{*^-}t^{\sigma}} =0,\label{CRK2eqn}\\
\lefteqn{ \braket{\Delta^{-}s^{\sigma}}{V_{K^{0}}}{\Sigma^{*^-}s^{\sigma}}   \braket{\Sigma^{*^-}s^{\sigma}} {j_{em}^{\mu}} {\Sigma^{*^-}t^{\sigma}} }  \nonumber \\
& & - \braket{\Delta^{-}s^{\sigma}}{j_{em}^{\mu}}{\Delta^{-}t^{\sigma}}   \braket{\Delta^{-}t^{\sigma}} {V_{K^{0}}} {\Sigma^{*^-}t^{\sigma}} =0. \label{CRK3eqn}
\end{eqnarray}

{\sloppy
For each of the baryon pairs considered previously, Eqs.~(\ref{CRK1eqn})--(\ref{CRK3eqn}) hold whether or not $s_{x}$ is zero and imply that
}
\begin{eqnarray}
\braket{Bs^{\sigma},\lambda}{j_{em}^{\mu}(0)}{Bt^{\sigma},\lambda}=\braket{\Omega^{-}s^{\sigma},\lambda} {j_{em}^{\mu}(0)} {\Omega^{-}t^{\sigma},\lambda},
\label{CRMaineqn}
\end{eqnarray}

where $B= \Delta^{-}, \Xi^{*\,-}, \Sigma^{*\,-} ,\;\mbox{and } t_{z}\rightarrow \infty, \;  s_{z}\rightarrow \infty,\;\mbox{and}\;\lambda=\mbox{helicity}\;=+3/2.$

 Eq.~(\ref{CRMaineqn}) is obtained in \emph{broken} symmetry with $r\, (\mbox{constant})\,\geq 1$---thus ensuring no helicity reversal---and we evaluate it with $\mu=0$ and $0\leq \theta < \pi/2$ and $\sin \theta =s_{x}/\abs{\vec{s}}$ where $0\leq s_{x} < \abs{\vec{s}}$ using Eqs.~(\ref{matrixeqn})--(\ref{momvecp2eqn}).  We now obtain (see Eq.~(\ref{f2termvanishes}) below):

\begin{eqnarray}
\lefteqn{
\lim_{ \stackrel{s_{x}  \rightarrow 0}{    \stackrel{t_{z}\rightarrow +\infty}{\SSt s_{z}\rightarrow \SSt +\infty} }  }           \{    \cosh\left[\frac{\alpha_{B} -\beta_{B} }{2}\right] F_1^{B}(q^{2}_{B}) -2 F_2^{B}(q^{2}_{B}) (      \sinh \frac{\alpha_{B} }{2} \sinh \frac{\beta_{B} }{2} (\cosh
   \beta_{B} +1) \sin ^2 \frac{\theta }{2}      \}   } \hspace{7in}  \nonumber\\
=\lim_{ \stackrel{s_{x}  \rightarrow 0}{    \stackrel{t_{z}\rightarrow +\infty}{\SSt s_{z}\rightarrow \SSt +\infty} }  }           \{    \cosh\left[\frac{\alpha_{\Omega^{-}} -\beta_{\Omega^{-}} }{2}\right] F_1^{\Omega^{-}}(q^{2}_{\Omega^{-}}) - 2 F_2^{\Omega^{-}}(q^{2}_{\Omega^{-}}) (    \sinh \frac{\alpha_{\Omega^{-}} }{2} \sinh \frac{\beta_{\Omega^{-}} }{2} (\cosh
   \beta_{\Omega^{-}} +1) \sin ^2 \frac{\theta }{2}    \}  \hspace{0.4in}   \nonumber\\
\label{CRMain1eqn}
\end{eqnarray}

Taking the limits [$s_{x}  \rightarrow 0, t_{z}\rightarrow \infty, s_{z}\rightarrow \infty$] (the term $(1+r)/(2 \sqrt{r})$ is common to both sides of Eq.~(\ref{CRMain1eqn}) and can be cancelled and the coefficient of $F_2^{B}(q^{2}_{B})$ vanishes---see Eq.~(\ref{f2termvanishes}) below) in Eq.~(\ref{CRMain1eqn}) with $s_{z}=r t_{z}$ [$r\, (\mbox{constant})\,\geq 1$] then yields:
\begin{equation}\label{CRMain2eqn}
F_1^{B}(q^{2}_{B})= F_1^{\Omega^{-}}(q^{2}_{\Omega^{-}}).
\end{equation}

Eq.~(\ref{CRMain2eqn}) for the electric charge form factor as a function of four-momentum transfer squared is the main result of this work. We note that in deriving  Eq.~(\ref{CRMain2eqn}), even though $\abs{\vec{s}}\,\mbox{and}\,\abs{\vec{t}}\rightarrow +\infty$, $q^{2}_{B}$ is finite and $q^{2}_{B} = -\frac{(1-r)^2}{r}m^2_{B}-\frac{s_{x}^2}{r}\equiv -Q^{2}_{B}$.
\begin{eqnarray}
  {q^{2}_{B}}_{\mid s_{x}\rightarrow 0} &=& -\frac{(1-r)^2}{r}m^2_{B}\,,
  \label{qsqcollineareqn}
\end{eqnarray}
\begin{eqnarray}
  \cosh\left[\frac{\alpha_{B} -\beta_{B} }{2}\right] &\rightarrow& \frac{1+r}{2 \sqrt{r}}\,, \nonumber\\
  \sinh \frac{\alpha_{B} }{2} \sinh \frac{\beta_{B} }{2} (\cosh
   \beta_{B} +1) \sin ^2 \frac{\theta }{2}  &\stackrel{s_{x}\rightarrow 0}{\rightarrow}&  0\,,
   \label{f2termvanishes}
\end{eqnarray}
where $B=\Delta^{-}\mbox{, }\Sigma^{*^-}  \mbox{, }\Xi^{*^-} \mbox{, or }\, \Omega^{-}$.

Explicitly, Eq.~(\ref{CRMain2eqn}) reads:
\begin{eqnarray}\label{CRMain3eqn}
F_1^{\Delta^{-}}(q^{2}_{\Delta^{-}}) = F_1^{\Sigma^{*^-}}(q^{2}_{\Sigma^{*^-}}) = F_1^{\Xi^{*^-}}(q^{2}_{\Xi^{*^-}}) =
 F_1^{\Omega^{-}}(q^{2}_{\Omega^{-}}) \quad  \mbox{and }\nonumber\\
F_1^{\Delta^{-}}(0) =  F_1^{\Sigma^{*^-}}(0)    = F_1^{\Xi^{*^-}}(0)        =F_{1}^{\Omega^{-}}(0)=-1\, ,\nonumber\\
\mbox{since } r=1 \Rightarrow q^{2}_{\Delta^{-}}=q^{2}_{\Sigma^{*^-}}   =  q^{2}_{\Xi^{*^-}}  =  q^{2}_{\Omega^{-}}=0.
\end{eqnarray}



We have shown that electric charge form factor $F_1^{B}(q^{2}_{B})$ for the ground state \emph{physical} decuplet $U$-spin $=\frac{3}{2}$ quartet members ($B=\Delta^{-}\mbox{, }\Sigma^{*^-}  \mbox{, }\Xi^{*^-} \mbox{, or } \: \Omega^{-}$) are \emph{analytically} the same without ascribing any specific form to their quark structure or intra-quark interactions or assuming an effective Lagrangian.  They differ only in that they are functions of mass dependent four-momentum transfer $q^{2}_{B}$.  It is clear that future experimental measurements of the $\Omega^{-}$ magnetic moment and accessible form factors for $q^{2}_{\Omega^{-}}\leq 0$, while very difficult, will certainly have great importance for viable theoretical models (especially lattice QCD models) of the structure of baryons.

\newpage

\bibliographystyle{apsrev}

\end{document}